\documentclass[11pt,twocolumn,letterpaper]{article}
\usepackage[margin=0.72in,columnsep=0.25in]{geometry}
\usepackage[T1]{fontenc}
\usepackage[utf8]{inputenc}
\usepackage{mathptmx,tgtermes,helvet}

\usepackage{microtype,graphicx,booktabs,amsmath,amssymb,xcolor,stfloats}
\usepackage{natbib,caption,longtable,array,balance,placeins}
\definecolor{PaperBlue}{HTML}{4F587D}
\usepackage[colorlinks=true,allcolors=PaperBlue]{hyperref}
\newcommand{\CorpusCount}{564,861}
\newcommand{\HistoricalCount}{93,165}

\newcommand{\LastSubmission}{2026-10-27}

\hypersetup{pdftitle={"PRISM" Is Not All You Need: Cross-Topic Name Collisions and Scientific Paper Retrieval},pdfauthor={Jingjie Ning, Xueqi Li, Yibo Kong},pdfkeywords={scientific naming, cross-topic convergence, name collisions, paper retrieval, acronym disambiguation, language-model naming}}

\newcommand{\C}{\mathcal{C}}

\begin{document}
\AddToHookNext{shipout/foreground}{\put(28,-8){\raisebox{-\height}{\includegraphics[width=1.40in]{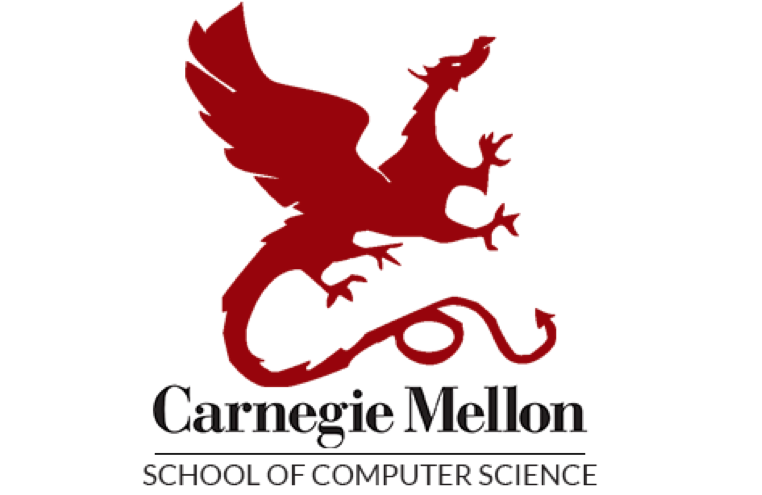}}}}
\twocolumn[{
\begin{center}
{\LARGE\bfseries ``PRISM'' Is Not All You Need:\\[3pt]
Cross-Topic Name Collisions and Scientific Paper Retrieval\par}
\vspace{7pt}
{\large Jingjie Ning\textsuperscript{*}\quad Xueqi Li\quad Yibo Kong\par}
\vspace{3pt}
{\normalsize School of Computer Science, Carnegie Mellon University\par}
\vspace{2pt}
{\small\{\href{mailto:jening@cs.cmu.edu}{jening}, \href{mailto:xueqil@cs.cmu.edu}{xueqil}, \href{mailto:yibok@cs.cmu.edu}{yibok}\}@cs.cmu.edu\par}
\vspace{5pt}
\end{center}
\begin{center}
\begin{minipage}{0.85\textwidth}
\small
\centerline{\textbf{Abstract}}
\vspace{4pt}
Scientific names increasingly connect papers from different research topics, leaving readers to recover a paper's identity from context. We study this convergence in \CorpusCount{} AI-related arXiv records. Cross-topic name collision grows 4.33-fold from a pooled 2020--2022 reference to 2026, compared with 1.12-fold within topics. A second sentence encoder reproduces the pattern, and name-group resampling quantifies its uncertainty. Fixed word-frequency bands locate the main increase in an intermediate-frequency vocabulary shared by many research areas. We then measure the ambiguity that remains after supplying research clues. With the same three-term query construction, shared-name targets retain matched identification deficits of 25.7 percentage points under title-based BM25 and 6.9 points under title-plus-abstract E5. Within fixed same-name candidates, listwise language-model ranking reaches 89.2\% using the query and candidate titles. Topic-specific comparisons show that context resolves more cross-topic ambiguity. The evaluation separates candidate competition, document fields, and contextual ranking through lexical, dense, and language-model baselines. Source-linked cases and a portable evaluation collection make the phenomenon inspectable and reusable for scientific retrieval and research assistants.
\end{minipage}
\end{center}\vspace{5pt}
\begin{minipage}{\textwidth}
\centering\includegraphics{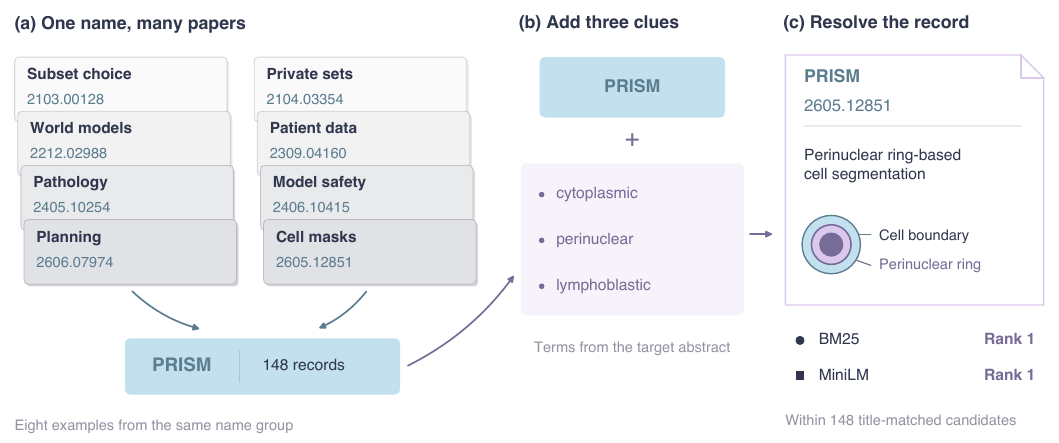}
\captionof{figure}{\textbf{Research clues identify a specific paper behind a shared name.} Eight PRISM examples lead to one candidate group; three abstract-derived terms identify the target under both title rankers. The cell icon is schematic.}
\label{fig:teaser}
\end{minipage}\vspace{8pt}
}]
\begingroup
\renewcommand{\thefootnote}{\fnsymbol{footnote}}
\footnotetext[1]{Corresponding author.}
\endgroup

\section{Introduction}

PRISM can name a spatial world model, a pathology encoder, or a safety method. TRACE can reconstruct tables or attribute model outputs. SAGE can denote surgical mapping or audio encoding. A memorable label helps readers retain a name, while the surrounding research context determines which paper that name denotes.

Scientific names increasingly connect different research topics. We call a shared exact label across distinct bibliographic records a \emph{name collision}. Cross-topic convergence is an increasing collision rate among records assigned to different semantic topics. This distinction separates reuse within a research area from a label that links unrelated parts of the literature. It also poses an information-retrieval question. How precisely does a name identify a paper, and which clues resolve ambiguity?

Our corpus covers \CorpusCount{} AI-related arXiv papers submitted from 2020 through \LastSubmission{}, with complete original-version metadata and a 2015--2019 extension. Cross-topic collision grows 4.33-fold over the pooled 2020--2022 reference, while the within-topic point estimate grows 1.12-fold. The increase survives author-overlap filters and changes in topic resolution. A second sentence encoder tests the dependence on the primary representation. Name-group resampling quantifies the influence of the observed label vocabulary.

The affected population is substantial. In the fixed retrieval collection, 28,152 papers, or 19.5\% of named records, have a namesake. A shared name gives a candidate set; a useful retrieval system must identify the intended member. We therefore measure the residual difference after both shared-name and metadata-matched unique-name targets receive three terms from their source abstracts. Full-index BM25 reaches 59.4\% and 85.1\%, respectively, leaving a 25.7-point matched gap. This comparison supplies a practical measure of retrieval difficulty under the same query-construction rule.

Specific research context restores much of the identifying information. Exact-name candidate retrieval fixes the competing records, and the three terms raise BM25 Top-1 from 30.8\% to 70.6\%. Relative error reductions and two-candidate comparisons locate a larger benefit in resolving competitors from different topics. A source-linked pair of AutoInit papers illustrates the residual difficulty when a name and a research goal are both shared. Dense retrieval and language-model reranking broaden the evaluation across scoring approaches.

Word-frequency bands reveal where the historical increase accumulates. The analysis keeps each label's frequency fixed across years and separates its prevalence from repetition within its frequency band. Fixed bands expose how concentration varies across the reference vocabulary. Frequent names span many abstract-derived topics, while the spelling and source evidence distinguish exact-name ambiguity from reuse within a named method lineage.

We contribute a version-aware measurement of scientific-name convergence, a retrieval evaluation separating name specificity from contextual disambiguation, and a portable evaluation collection with a local query demonstration. The companion artifact contains the fixed corpus, candidate groups, queries, topic-error labels, baselines, and evaluation entry point. The appendix provides the controlled model-naming case.

\section{Related work}
\label{sec:related}

\paragraph{Scientific naming and acronym ambiguity.}
\citet{agra} organize academic abbreviation through analysis, generation, and ranking. \citet{acronym_growth} document increasing acronym use in biomedical literature, and \citet{method_names} discuss memorability, uniqueness, and checking existing names. Scientific-document benchmarks study acronym identification and disambiguation \citep{acronym_tasks}. We measure topic spread and retrieval specificity \mbox{of shared labels.}

\paragraph{Shared lexical preferences.}
The One-Word Census measures answer-choice concentration across models \citep{census}. The Ghost Couple studies correlated personal-name priors \citep{ghost}, and brand experiments separate spontaneous mentions from recommendations \citep{taste}. Artificial Hivemind analyzes repeated outputs and cross-model homogeneity \citep{hivemind}; subsequent work emphasizes reference distributions \citep{hivemind_review}. Verbalized Sampling studies generation diversity \citep{verbalized}, while Linguistic Monoculture considers shared assistance at population scale \citep{monoculture}. Our lexical strata and supporting model case make the reference vocabulary explicit.

\paragraph{Scientific writing and retrieval.}
\citet{llm_scientific_writing} estimate LLM-modified scientific writing with population-level word distributions. We reuse their published reference weights for an exploratory writing-style association, with original abstracts and within-year topic controls. BM25 supplies a lexical retrieval reference \citep{bm25}; sentence encoders support semantic ranking and topic partitioning \citep{sentence_bert}. UMAP provides a qualitative projection of the name-bearing records \citep{umap}. E5 adds contrastively trained text embeddings \citep{e5}. The retrieval comparisons fix paper identities and specify the candidates, indexed evidence, and queries.

\begin{figure*}[t]
\centering\includegraphics{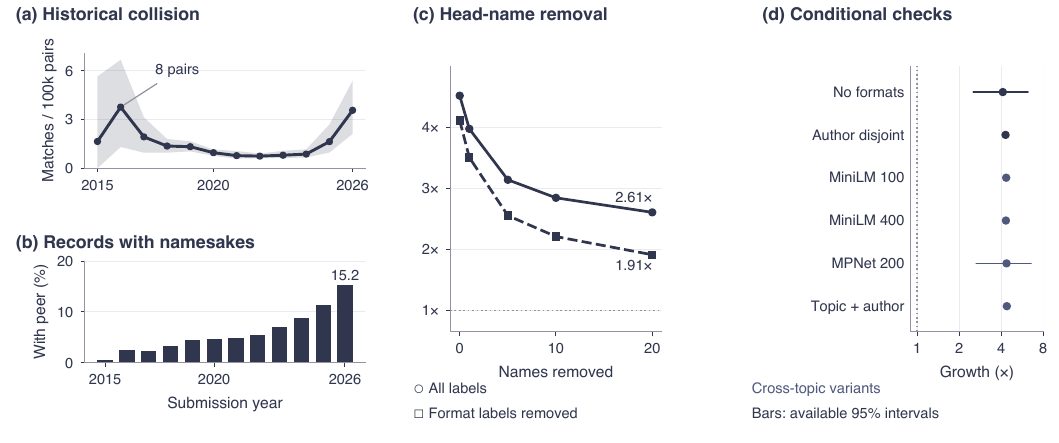}
\caption{\textbf{Name reuse increases across extraction and population definitions.} Historical intervals resample name groups. The robustness panel labels each estimand and uses the pooled 2020--2022 reference; head removal is shown separately. The 2026 cohort ends on October 27.}
\label{fig:namespace}
\end{figure*}

\section{Cross-topic name convergence}
\label{sec:corpus}

\subsection{Population, labels, and uncertainty}

Each collected arXiv identifier enters once when it carries a cs.AI, cs.LG, cs.CL, cs.CV, cs.IR, cs.RO, cs.MA, or stat.ML tag. Cornell's version-309 metadata snapshot \citep{arxiv_dataset} and official updates supply titles, abstracts, authors, dates, and version histories. All primary records retain original titles and abstracts; \HistoricalCount{} earlier eligible records extend the trajectory to 2015.

A \emph{title label} is a compact leading token followed by a colon, such as \texttt{PRISM: \ldots}. Unicode normalization and case folding define its comparison key. Suffixes remain part of the key. A \emph{method lineage} consists of papers developing or reusing one named method. Our population counts concern record identities; source-linked cases provide evidence about their research objects and lineages.

For cohort $t$, let $n_t(a)$ count records with label $a$ and $N_t=\sum_a n_t(a)$. The probability that two distinct labelled records match is
\begin{equation}
\C_t=\frac{\sum_a n_t(a)(n_t(a)-1)}{N_t(N_t-1)}.
\label{eq:corpus_collision}
\end{equation}
The number of matching unordered pairs is ${N_t\choose2}\C_t$. The mean number of same-name peers per labelled record is
\begin{equation}
B_t=(N_t-1)\C_t.
\label{eq:record_burden}
\end{equation}
We report collision per 100,000 eligible pairs. The reference pools 2020--2022 records, including cross-year pairs; an alternative pools within-year pair counts.

We report 95\% percentile intervals from 2,000 paired draws of complete name groups. A selected name group retains its records and topic composition. Duplicate draws are separate resampled groups, so matching-pair counts scale linearly with group multiplicity. Eligible-pair denominators are recomputed from the resampled records. The same multiplicities apply across periods. These intervals quantify sensitivity to the observed name-group population. Appendix~\ref{app:measurement} defines \mbox{extraction and resampling.}

Topic-conditioned rates separate name reuse from the number of available paper pairs. Let $n_{ak}$ count records with name $a$ in topic $k$, and let $N_k$ count all labelled records in that topic, with $N=\sum_k N_k$. The cross-topic rate is
\begin{equation}
\C_{\mathrm{cross}}=
\frac{\sum_a\left[n_a(n_a-1)-\sum_k n_{ak}(n_{ak}-1)\right]}
{N(N-1)-\sum_k N_k(N_k-1)}.
\label{eq:cross_topic}
\end{equation}
The within-topic rate uses the two subtracted sums as numerator and denominator. A growing topic changes both its available pairs and its matching pairs. Conditioning on the relation makes their ratio comparable across years. Cross-topic and within-topic rates describe different eligible populations; their trajectories compare each rate with its own pooled reference.

\subsection{Cross-topic growth}

The 2026 title-label collision rate is 4.52 times the pooled reference, with a name-group interval of [2.59, 7.09]. The complete historical trajectory includes the early small cohorts, including 652 labelled records and eight matching pairs in 2016. Recent growth is accompanied by greater exposure to namesakes. In 2026, 15.2\% of labelled records have at least one same-year peer, and the mean is 1.55 peers per labelled record (Figure~\ref{fig:namespace}).

\begin{figure*}[t]
\centering\includegraphics{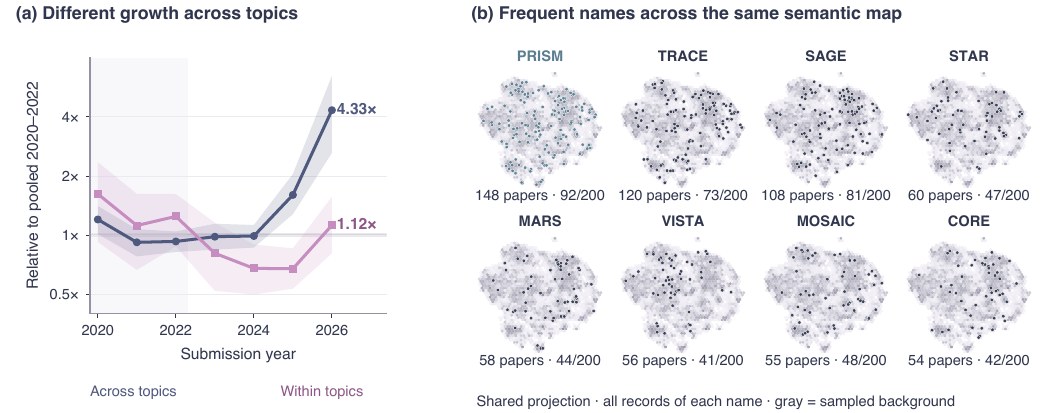}
\caption{\textbf{Cross-topic collisions grow as frequent names span the semantic map.} Indexed rates use their corresponding eligible pairs. The eight most frequent names share one projection and background; annotations give record and occupied-cluster counts. Topic statistics use the original abstract vectors.}
\label{fig:topics}
\end{figure*}

Cross-topic collision grows much faster than the within-topic point estimate. In 144,523 non-format-labelled records, a 200-cluster partition gives cross-topic growth of 4.33 [2.62, 6.47] and within-topic growth of 1.12 [0.81, 1.57]. The partitions use abstracts with their own title labels masked. Using 100 or 400 clusters gives cross-topic point ratios from 4.33 to 4.35. Figure~\ref{fig:topics} indexes each trajectory to its own pooled reference so the relative change is directly visible.

The increase spans records with different listed authors. Excluding cross-topic pairs that share a normalized author name preserves 4.40-fold growth. Author-only filtering under the common reference gives 4.30-fold growth. Removing six publication-format labels and the 20 most frequent remaining names leaves a 1.91-fold increase [1.66, 2.18]. First-version, case-sensitive, and fixed-category-mixture analyses retain the recent upward direction.

The second encoder reproduces both historical growth and the harder same-topic retrieval pattern. Re-encoding all 144,523 masked abstracts with MPNet gives cross-topic growth of 4.38 [2.65, 6.56] and within-topic growth of 1.14 [0.77, 1.67]. Under this partition, the two-candidate retrieval comparison contains 469 cross-topic and 71 same-topic targets. Contextual BM25 identifies 80.9\% and 63.4\%, respectively, a difference of 17.5 points [6.3, 28.4]. The original dense ranker gives a corresponding difference of 18.7 points [7.9, 29.9]. The partition changes while the queries, candidates, and ranking outputs remain fixed. Thus, the observed distinction extends across two separately trained topic representations.

\subsection{A shared lexical vocabulary}

Intermediate-frequency names account for most of the observed increase. We assign each alphabetic label a fixed positive English token frequency $f$ when the resource attests that label from wordfreq 3.1.1 \citep{wordfreq} and define its Zipf frequency as $z=\log_{10}f+9$. A unit increase means a tenfold increase in reference frequency. Unattested and nonalphabetic labels form a separate stratum. Positive-frequency labels enter five fixed bands with boundaries at 2, 3, 4, and 5. These bands describe lexical familiarity in the reference resource, which includes abbreviations and proper names.

Names with $3\leq z<5$, corresponding to one to fewer than 100 occurrences per million tokens, account for 12.4\% of 2026 named records in the fixed collection and 89.4\% of cross-topic matching pairs. Together they contribute 98.0\% of the net increase in the population cross-topic collision rate. The two bands have conditional growth of 4.98 [2.72, 7.82] and 4.42 [2.25, 7.64]. The next lower band, $2\leq z<3$, grows 2.41-fold [1.76, 3.32], and the highest-frequency band grows 2.51-fold [1.74, 3.58]. Unattested and nonalphabetic labels have a ratio of 0.56 [0.46, 0.67]. The intermediate bands combine high collision mass with larger growth point estimates. The intervals resample complete observed name groups.

\begin{figure*}[t]
\centering\includegraphics{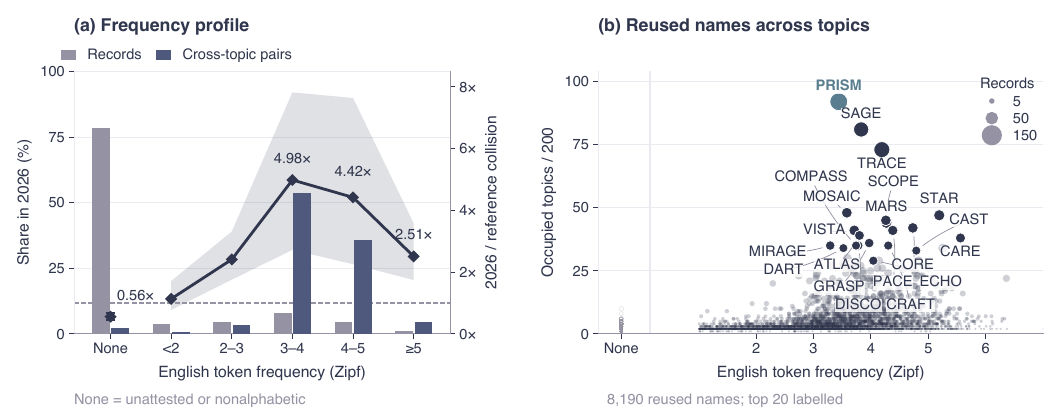}
\caption{\textbf{Intermediate-frequency names carry most of the cross-topic increase.} Fixed Zipf bands separate record share, matching-pair share, and conditional growth. Name-level topic breadth uses the common 200-cluster partition, with point area encoding records and labels chosen by frequency.}
\label{fig:lexical_growth}
\end{figure*}

Each band's conditional rate uses different-topic pairs drawn entirely from that band. Its contribution to the population rate divides its cross-topic matching pairs by the full population's cross-topic denominator. The contributions sum exactly because matching labels always occupy the same fixed band. Figure~\ref{fig:lexical_growth} therefore separates a band's prevalence, concentration, and contribution to growth. The scatter shows how record frequency and topic coverage vary across individual names; its 20 labelled points are selected by full-corpus record count. The exported measurements retain every band and name, including high-frequency and unattested labels.

\begin{figure*}[t]
\centering\includegraphics{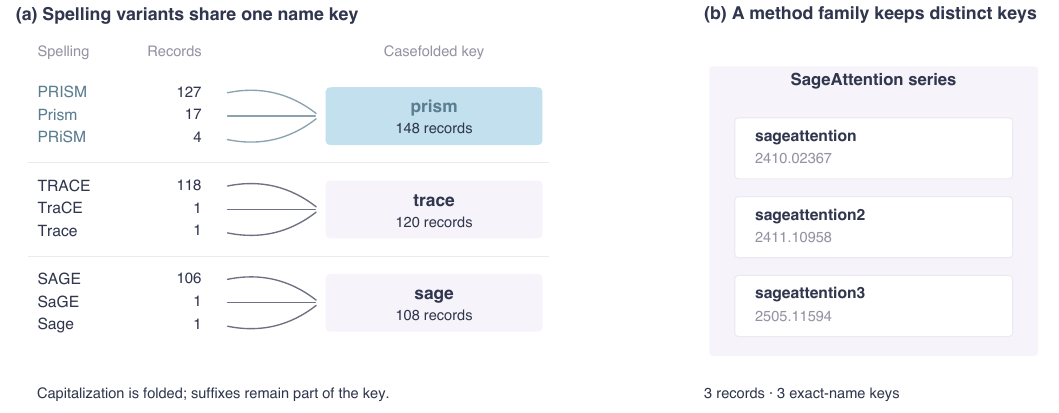}
\caption{\textbf{Spelling equivalence and method lineage define different relationships.} Case folding merges capitalization variants into one name group. Suffixes retain distinct keys, so the three source-linked SageAttention papers have separate retrieval labels.}
\label{fig:cases_main}
\end{figure*}

\subsection{Names connect distinct research objects}

The frequent labels connect several research areas. PRISM, TRACE, and SAGE occupy 92, 73, and 81 of the 200 semantic clusters across 148, 120, and 108 records. Figure~\ref{fig:topics} uses a common projection for the eight most frequent groups, and Figure~\ref{fig:cases_main} separates spelling equivalence from source-linked method families. Raw topic coverage describes the observed uses of each name. A fixed-volume analysis of the 18 names with at least five uses in both periods yields 4.92 and 4.83 expected topics in a sample of five records, with a paired difference interval of [$-0.21$, 0.03]. This locates the growth in greater reuse of a vocabulary whose recurring members already cover many topics.

\paragraph{PRISM links different research objects through flexible spelling.}
The 148 exact casefolded records include 127 \texttt{PRISM}, 17 \texttt{Prism}, and four \texttt{PRiSM} titles. Parameterized submodular information measures select letters inside \emph{PaRameterIzed} \citep{prism_submodular}; spatial world models use probabilistic real-time inference \citep{prism_world}; model safety uses private, robust, independent safety measures \citep{prism_safety}. These sources describe different research objects behind the common string. Capitalization therefore changes the measured group size while preserving substantial reuse. In 144 of the 148 records, PRISM already appears \mbox{at initial submission.}

\paragraph{TRACE links several kinds of reconstruction and attribution.}
Its uses include table reconstruction \citep{trace_table}, attribution of model outputs to sources \citep{trace_attribution}, time-series adaptation \citep{trace_timeseries}, and agent-trajectory evaluation \citep{trace_evaluation}. These uses attach a familiar process word to different tasks and research objects. The current group contains 120 records, including 118 uppercase spellings, and 116 retain the name from their first title. The high uppercase share makes TRACE's spelling sensitivity small. Its broad topic distribution and the source descriptions together establish that readers encounter multiple task-specific meanings under the same written query.

\paragraph{SAGE distinguishes exact-name ambiguity from a named lineage.}
Endoscopic mapping \citep{sage_slam}, Semantic Graph Entropy written \texttt{SaGE} \citep{sage_entropy}, and audio encoding \citep{sage_audio} supply distinct uses. The 108-record casefolded group includes 106 uppercase titles. Case folding includes the mixed-case entropy name, while the extraction rule preserves suffixes in their own keys. SageAttention and its numbered successors therefore remain separate labels, with their relationship recorded as a source-declared lineage \citep{sageattention,sageattention2,sageattention3}. This example makes the unit of analysis concrete. Exact-name matching defines the records competing for a short query; source evidence identifies relationships among the underlying methods. Appendix~\ref{app:cases} supplies the complete source-linked cases and variant counts.

\section{Names lose retrieval specificity}
\label{sec:retrieval}

\subsection{Residual ambiguity after research context}

Shared-name targets remain harder to identify after receiving the same type of research clue. We search the fixed 144,523-title collection for 1,400 shared-name targets and 1,400 unique-name targets matched without replacement on year, category, and title-length interval. The context condition appends three terms selected from each target abstract by one corpus-wide rule. Full-index BM25 Top-1 is 85.1\% for unique-name targets and 59.4\% for shared-name targets. The matched difference is 25.7 percentage points [22.5, 28.7]. MiniLM gives a corresponding difference of 13.2 points [9.8, 16.6]. These contrasts describe residual identification difficulty across matched populations.

\begin{figure*}[t]
\centering\includegraphics{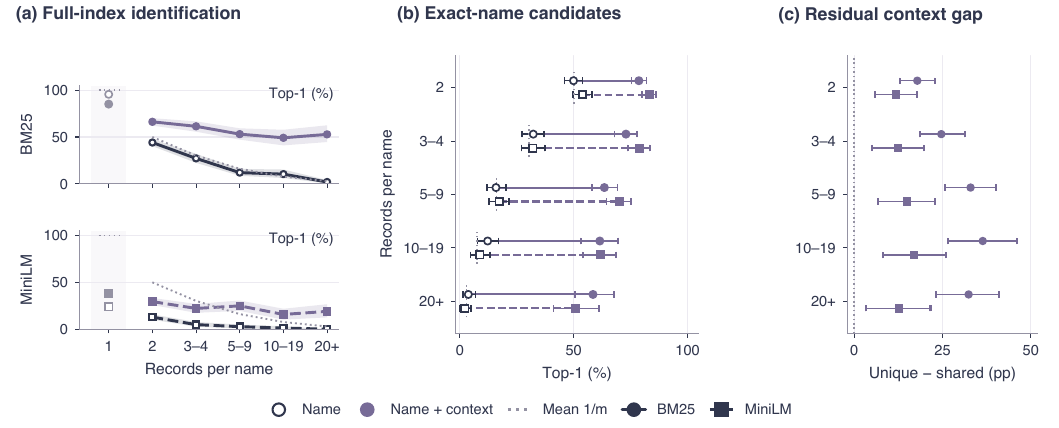}
\caption{\textbf{Research clues recover many shared-name targets and leave a measurable residual gap.} Full-index results, fixed-candidate gains, and contextual gaps use the same targets. Gray references average $1/m$ across actual groups; intervals resample complete shared-name groups and their matched controls.}
\label{fig:retrieval_specificity}
\label{fig:retrieval}
\end{figure*}

The contextual difference extends across group sizes. Shared names with two records reach 66.2\% BM25 Top-1, and names with at least 20 records reach 52.8\%. Figure~\ref{fig:retrieval_specificity} compares each group with its own matched unique-name controls. The matches preserve observable metadata; the name groups retain their actual contents and lexical characteristics. The contrast answers how well the same clue-generation procedure identifies papers in these populations. Section~\ref{sec:strong_rankers} repeats the comparison with E5 and adds the abstracts to the indexed evidence.

\paragraph{The structural reference.}
A name alone supplies a shared query for all $m$ records in its group. If the intended target is uniformly distributed over those records, a target-independent first-rank choice within that group has mean accuracy $1/m$. This statement holds for a deterministic winner as well as a uniformly resolved tie. BM25 title-length normalization can give namesakes different scores, and full-index retrieval can rank other labels above them. The reference therefore calibrates the identity information in a name. The measured contextual gaps and candidate-ranking gains establish how much specific evidence resolves that ambiguity.

The affected population extends beyond the leading labels. Across the collection, 28,152 papers have a namesake, and an affected paper has a mean of 8.33 additional same-name peers. Full-index retrieval includes both same-name and differently named distractors. Exact-name candidates isolate selection within the correct name group.

\subsection{Specific research context recovers identity}

Context recovers substantial accuracy when the competing referents are held fixed. Exact-name candidate retrieval keeps every record whose normalized title label matches the query. BM25 retains full-index term statistics, and the dense encoder retains its title and query vectors. Three abstract-derived terms raise BM25 Top-1 from 30.8\% to 70.6\%, a 39.8-point gain [37.2, 42.3]. Dense Top-1 rises from 32.0\% to 74.0\%, a 42.0-point gain [39.1, 44.8].

The gains remain substantial for larger groups. With at least 20 candidates, contextual BM25 and dense Top-1 reach 58.5\% and 50.7\%, compared with 2.9\% for uniform candidate choice. A PRISM example makes the query information concrete. The terms \emph{cytoplasmic}, \emph{perinuclear}, and \emph{lymphoblastic} identify a cell-segmentation paper \citep{prism_cell} at rank one under both scoring functions within its 148-record name group.

The queries are automatically constructed from source abstracts. They measure identification when specific source clues are supplied. The benchmark preserves the query generator and supports additional reader-authored queries under the same target and candidate definitions.

\subsection{Harder ambiguity within a topic}
\label{sec:topic_resolution}

Every first-ranked outcome has an explicit paper identity. Let $A$ denote the probability of selecting the target, $E_{\mathrm{same}}$ the probability of selecting a different paper in its topic, and $E_{\mathrm{cross}}$ the probability of selecting a paper in another topic. Tied first ranks allocate equal probability to their records, giving $A+E_{\mathrm{same}}+E_{\mathrm{cross}}=1$ for each query. Adding context changes accuracy by the sum of the reductions in the two error categories. For category $g$, we report
\begin{equation}
R_g=\frac{E_g(\text{name})-E_g(\text{name+context})}
{E_g(\text{name})}.
\end{equation}
This relative error reduction answers how much of each initial ambiguity context resolves. It also separates the effect of context from the unequal starting prevalence of the two error categories.

Context removes a larger fraction of cross-topic error than same-topic error. Relative to the name-only error mass, BM25 removes 58.9\% [55.7, 62.4] of cross-topic error and 24.4\% [6.6, 40.0] of same-topic error. MiniLM removes 64.0\% [60.5, 67.5] of cross-topic error. Its same-topic reduction is 18.3\% [$-7.8$, 39.8], with \mbox{the interval spanning zero.}

\begin{figure*}[t]
\centering\includegraphics{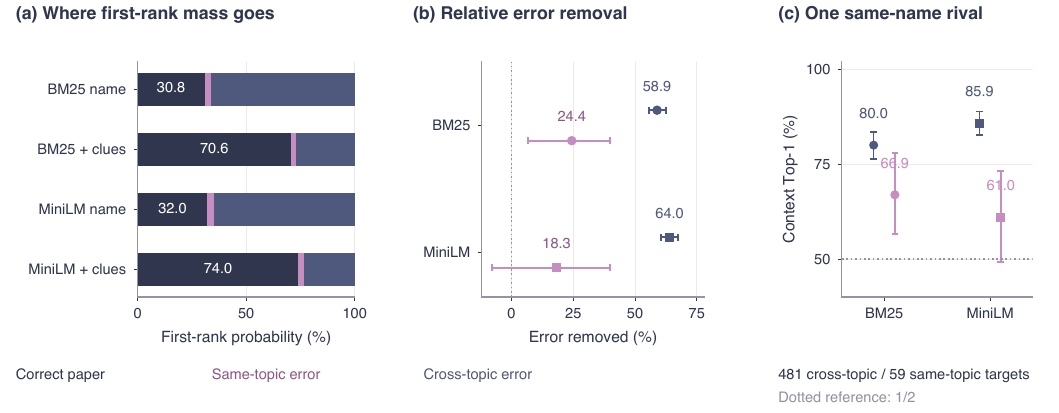}
\caption{\textbf{Context resolves a larger share of cross-topic error.} Stacked outcomes partition first-rank probability. Relative error reduction conditions on each initial error mass; the two-candidate comparison fixes the number of choices. Targets and candidates remain fixed.}
\label{fig:topic_resolution}
\end{figure*}

The fixed-size comparison reaches the same substantive distinction. Among two-record name groups, contextual BM25 identifies 80.0\% of targets with a cross-topic rival and 66.9\% with a same-topic rival. Dense ranking reaches 85.9\% and 61.0\%. The between-group differences are 13.1 points [1.6, 25.0] and 24.8 points [13.1, 37.0]. Holding candidate count at two removes the direct change in the number of choices and exposes a more specific distinction between research objects that already share both a name and a topic.

\paragraph{A same-topic residual case.}
The query \emph{AutoInit exploding initialization vanishing} targets analytic signal-preserving weight initialization \citep{autoinit_signal}. Its sole namesake initializes networks through Jacobian tuning \citep{autoinit_jacobian}. BM25, MiniLM, and E5 title ranking select the latter. DeepSeek listwise reranking recovers the intended record from the same two candidate titles. E5 also recovers it when the index includes the abstracts. The query describes the shared initialization goal, while the source papers distinguish signal-moment propagation from Jacobian-based tuning. We select this example as the earliest identifier among the ten two-candidate, same-topic targets missed by contextual BM25 and MiniLM. The contrast illustrates how ranking functions use a clue that describes a shared research goal. The exported case preserves the query, candidate titles, topics, and complete ranking results.

\subsection{Retrieval fields and stronger ranking functions}
\label{sec:strong_rankers}
A stronger dense encoder preserves the distinction between finding a paper in the collection and selecting it among namesakes. E5-base-v2 \citep{e5} searches the same titles with its documented query and passage prefixes. On the shared-name context queries, it reaches 46.6\% Top-1 in the full collection and 72.8\% inside exact-name candidates. The corresponding BM25 rates are 59.4\% and 70.6\%, and MiniLM gives 24.4\% and 74.0\%. Candidate restriction changes the competition seen by the ranking function while preserving each target and query. Table~\ref{tab:baselines} reports all three ranking metrics on the common populations.

Listwise language-model ranking resolves more of the fixed candidate ambiguity under the tested configuration. We ask DeepSeek-Flash to rank every candidate title for each of the same 1,400 shared-name context queries. A target-independent seeded permutation fixes candidate order. The output must contain each supplied candidate identifier exactly once. Across all assigned queries, Top-1 is 89.2\% [87.5, 90.9], Top-10 is 98.9\%, and mean reciprocal rank is 0.933. The two invalid outputs receive zero credit for every metric. The system selects among source records supplied in the prompt; the evaluation retains its complete ordering and response for each query.

The comparison keeps query evidence and document fields explicit. All title-only rows expose the same candidate text. Full-index search considers every indexed title, while exact-name ranking receives the complete observed namesake group. These scopes serve distinct stages of scientific search. A full-index retrieval score captures candidate discovery and target ranking together; exact-name ranking measures selection within a known group. The query demonstration exposes both the name group and the source record selected by research context.

Adding the source abstract improves identification while preserving a smaller shared-name gap. With the same E5 encoder and the same three-term queries, title-plus-abstract retrieval reaches 87.9\% Top-1 for shared-name targets and 94.8\% for matched unique-name targets. The paired residual difference is 6.9 points [4.8, 9.0], compared with 15.3 points [11.6, 18.9] for E5 over titles. Within exact-name candidates, the richer field reaches 93.6\% [92.3, 94.8]. The index now contains the abstract from which the query terms were extracted. This controlled field comparison measures how providing that evidence changes identification.

The two interventions address different parts of the search task. Candidate restriction specifies which label group the reader intends; richer document fields supply evidence for distinguishing its members. Keeping their outcomes separate exposes both contributions. In the title-plus-abstract condition, exact-name selection reaches 93.6\% against 87.9\% full-index accuracy on the identical shared-name targets. Under title-only evidence, listwise ranking reaches 89.2\% with the complete candidate group. These configurations provide concrete retrieval and reranking targets under explicit information budgets.

\FloatBarrier
\section{Reusable evaluation and implications}
\label{sec:reuse}

\begin{table*}[t]
\centering
\footnotesize
\setlength{\tabcolsep}{8pt}
\caption{\textbf{Contextual retrieval separates candidate scope and indexed evidence.} All rows use the same 1,400 shared-name targets. Top-1 intervals resample name groups; metrics use percentages. Title-plus-abstract rows expose the source of the query clues.}
\label{tab:baselines}
\begin{tabular}{llrrr}
\toprule
Ranker & Indexed fields & Top-1 [95\% interval] & Top-10 & MRR \\
\midrule
\multicolumn{5}{l}{\textit{Full collection, 144,523 records}} \\
BM25 & Title & 59.4 [56.8, 61.9] & 87.7 & 69.1 \\
MiniLM & Title & 24.4 [22.1, 26.7] & 43.9 & 31.2 \\
E5-base-v2 & Title & 46.6 [44.1, 49.4] & 68.7 & 54.3 \\
E5-base-v2 & Title + abstract & 87.9 [86.1, 89.5] & 97.4 & 91.6 \\
\midrule
\multicolumn{5}{l}{\textit{Exact-name candidates, complete observed groups}} \\
BM25 & Title & 70.6 [68.1, 73.0] & 96.7 & 80.6 \\
MiniLM & Title & 74.0 [71.4, 76.5] & 98.2 & 84.0 \\
E5-base-v2 & Title & 72.8 [70.3, 75.4] & 97.2 & 82.6 \\
DeepSeek-Flash & Title & 89.2 [87.5, 90.9] & 98.9 & 93.3 \\
E5-base-v2 & Title + abstract & 93.6 [92.3, 94.8] & 99.8 & 96.4 \\
\bottomrule
\end{tabular}
\end{table*}

The evaluation collection makes paper identity explicit at every stage. Its fixed corpus contains 144,523 labelled records, including 28,152 shared-name records in 8,190 name groups. The reference evaluation provides 1,400 shared-name targets and 1,400 metadata-matched unique-name targets, with name, field, and content-query variants. Tasks cover full-index retrieval and ranking within exact-name candidates. The evaluator scores Top-1, Top-10, and reciprocal rank, including explicit tie groups and missing submissions.

The companion artifact index links the corpus, candidates, queries, relevance records, baseline outputs, and topic-error labels. It includes an evaluation command, a data card, and a local query demonstration. The demonstration accepts a candidate name and an optional abstract, returns the source-linked namesakes, and ranks them by their semantic relation to the supplied text. The distribution includes this paper at a matching relative path so the artifact link remains usable after unpacking. The collection fixes the evaluation population, and the demonstration exposes its contents for inspection and reuse.

The target is a source-record identifier, so a related namesake counts as an identification error. Top-1, Top-10, and reciprocal rank use explicit tie groups. Omitted queries receive zero credit and contribute to the reported coverage. Each outcome links to its query, candidate identifiers, and topic assignments.

\paragraph{An executable identification task.}
Each input contains a query identifier, its supplied text, and the selected retrieval scope. The query identifier joins the target record to the frozen relevance file. For full-index search, systems rank the common collection. For exact-name selection, the supplied group contains every observed record with the normalized label. A prediction is an ordered list of source identifiers or an ordered sequence of explicit tie groups. This interface allows lexical scoring, vector search, and generative ranking to use the same evaluator. The source identifiers make the target and returned papers directly inspectable.

The query variants separate three amounts of supplied information. Name-only queries contain the observed title label. Field queries append the paper's assigned research category description. Context queries append three alphabetic terms selected by frequency and inverse document frequency over the abstract collection. The target label and a fixed stop list are removed from term selection. Each retained term has at least four characters, and the rule applies across the full corpus. Candidate-group uniqueness plays no role in choosing those words. All models and field extensions receive \mbox{the same query strings.}

\paragraph{Reproduction and extension.}
The package contains 60,200 scored outcomes with query identifiers, candidate scope, indexed fields, rank bounds, and outcome status. It retains the original lexical and MiniLM results together with the E5 and language-model extensions. The standalone BM25 and E5 entry points reconstruct candidate rankings from the exported corpus and queries. A separate command scores the retained language-model orderings. The E5 model revision and prefixes, language-model prompt and candidate permutations, and per-query failure states accompany their outputs. The package supports \mbox{rescoring and model reruns.}

Name groups provide the unit for paired uncertainty and for separating future development and evaluation data. When several targets share a name, they also share a candidate population. Resampling each target independently would discard that dependence. The reported intervals keep the group's targets and conditions together, and matched-control analyses keep the associated unique targets in the same draw. A new learned system can reserve complete groups for evaluation while using the remaining groups for development. The queries and outcomes are published; their group membership remains explicit.

\paragraph{For authors.}
Name checking is most useful before a label becomes the main retrieval key. The lexical analysis motivates checking familiar-word candidates across research topics, and the query results support publishing a short task descriptor together with the full title and a stable identifier. A descriptive suffix can distinguish related releases; its effectiveness should be evaluated with the intended search system.

\paragraph{For repositories and conferences.}
Submission interfaces can surface existing exact-name records and show their research contexts. The aim is an informative collision notice, with links and identifiers that let authors choose an appropriate label. Version-aware storage preserves when a name was introduced and how later releases relate to it.

\newpage
\paragraph{For search systems and research assistants.}
A shared name should expose a candidate set. Specific task or method context supports ranking within that set, and a full title plus stable identifier makes the selected result verifiable. When candidates occupy the same topic, the interface can ask for a distinguishing application, method component, or other remembered detail. The measured differences support keeping candidate retrieval and contextual selection separately inspectable.

\section{Conclusion}

Scientific names increasingly converge across research topics. The version-aware corpus records 4.33-fold growth in cross-topic collision against 1.12-fold growth within topics, with the pattern reproduced by a second sentence encoder. Fixed frequency bands locate most of the increase in intermediate-frequency names. Source-linked PRISM, TRACE, and SAGE cases connect this population pattern to distinct scientific objects, observed spellings, and documented method lineages.

Specific research clues recover many intended papers while leaving a measurable residual identification gap. Shared-name targets remain 25.7 points behind matched unique-name targets under title-based BM25 and 6.9 points behind under title-plus-abstract E5. Within complete same-name candidate sets, contextual ranking reaches 70.6\% with BM25, 89.2\% with a language model, and 93.6\% with E5 over titles and abstracts. Topic-specific errors and fixed-size comparisons identify same-topic competitors as the harder ambiguity. The shared corpus, frozen queries, complete candidate groups, and 60,200 recorded outcomes make these comparisons directly reusable. Together, the findings support scientific search interfaces that expose namesakes, use research context to rank them, and preserve a verifiable paper identifier in the final result.
\label{sec:body_end}

\clearpage
\nobalance
\begingroup
\linespread{1}\selectfont
\small
\raggedright
\bibliographystyle{plainnat}
\bibliography{references}

@misc{mpnet_model,
  author = {{Sentence Transformers}},
  title = {{all-mpnet-base-v2}},
  year = {n.d.},
  note = {Model card, accessed October 29, 2026. \href{https://huggingface.co/sentence-transformers/all-mpnet-base-v2}{Model repository}},
}

@misc{wordfreq,
  author = {Robyn Speer},
  title = {{wordfreq}},
  year = {2023},
  note = {Version 3.1.1. \href{https://pypi.org/project/wordfreq/3.1.1/}{Software release}},
}

@misc{minilm_model,
  author = {{Sentence Transformers}},
  title = {{all-MiniLM-L6-v2}},
  year = {n.d.},
  note = {Model card, accessed October 29, 2026. \href{https://huggingface.co/sentence-transformers/all-MiniLM-L6-v2}{Model repository}},
}

@inproceedings{agra,
  author = {Jianbing Zhang and Yixin Sun and Shujian Huang and Cam-Tu Nguyen and Xiaoliang Wang and Xinyu Dai and Jiajun Chen and Yang Yu},
  title = {{AGRA: An Analysis-Generation-Ranking Framework for Automatic Abbreviation from Paper Titles}},
  booktitle = {Proceedings of the Twenty-Sixth International Joint Conference on Artificial Intelligence},
  year = {2017},
  pages = {4221--4227},
  doi = {10.24963/ijcai.2017/590},
}

@article{acronym_growth,
  author = {Adrian Barnett and Zoe Doubleday},
  title = {{The growth of acronyms in the scientific literature}},
  journal = {eLife},
  year = {2020},
  volume = {9},
  pages = {e60080},
  doi = {10.7554/eLife.60080},
}

@article{method_names,
  author = {{Nature Methods}},
  title = {{What's in a method name?}},
  journal = {Nature Methods},
  year = {2024},
  volume = {21},
  pages = {923},
  doi = {10.1038/s41592-024-02323-5},
}

@article{bm25,
  author = {Stephen Robertson and Hugo Zaragoza},
  title = {{The Probabilistic Relevance Framework: BM25 and Beyond}},
  journal = {Foundations and Trends in Information Retrieval},
  year = {2009},
  volume = {4},
  number = {1--2},
  pages = {1--174},
  doi = {10.1561/1500000019},
}

@article{tfidf,
  author = {Gerard Salton and Christopher Buckley},
  title = {{Term-weighting approaches in automatic text retrieval}},
  journal = {Information Processing and Management},
  year = {1988},
  volume = {24},
  number = {5},
  pages = {513--523},
  doi = {10.1016/0306-4573(88)90021-0},
}

@article{holm,
  author = {Sture Holm},
  title = {{A Simple Sequentially Rejective Multiple Test Procedure}},
  journal = {Scandinavian Journal of Statistics},
  year = {1979},
  volume = {6},
  number = {2},
  pages = {65--70},
  note = {\href{https://www.jstor.org/stable/4615733}{JSTOR~stable~record}},
}

@misc{arxiv_dataset,
  author = {{Cornell University}},
  title = {{arXiv Dataset}},
  year = {2026},
  note = {Metadata snapshot, version~309,~October~2026},
  howpublished = {\href{https://www.kaggle.com/datasets/Cornell-University/arxiv}{Kaggle~arXiv~dataset}},
}

@misc{e5,
  author = {Liang Wang and Nan Yang and Xiaolong Huang and Binxing Jiao and Linjun Yang and Daxin Jiang and Rangan Majumder and Furu Wei},
  title = {{Text Embeddings by Weakly-Supervised Contrastive Pre-training}},
  year = {2022},
  eprint = {2212.03533},
  archivePrefix = {arXiv},
  note = {\href{https://arxiv.org/abs/2212.03533}{arXiv~preprint~arXiv:2212.03533}},
}

@misc{autoinit_signal,
  author = {Bingham, Garrett and Miikkulainen, Risto},
  title = {{AutoInit: Analytic Signal-Preserving Weight Initialization for Neural Networks}},
  year = {2021},
  eprint = {2109.08958},
  archivePrefix = {arXiv},
  note = {\href{https://arxiv.org/abs/2109.08958}{arXiv~preprint~arXiv:2109.08958}},
}

@misc{autoinit_jacobian,
  author = {He, Tianyu and Doshi, Darshil and Gromov, Andrey},
  title = {{AutoInit: Automatic Initialization via Jacobian Tuning}},
  year = {2022},
  eprint = {2206.13568},
  archivePrefix = {arXiv},
  note = {\href{https://arxiv.org/abs/2206.13568}{arXiv~preprint~arXiv:2206.13568}},
}

@misc{prism_2014,
  author = {Zhang, Ziming and Saligrama, Venkatesh},
  title = {{PRISM: Person Re-Identification via Structured Matching}},
  year = {2014},
  eprint = {1406.4444},
  archivePrefix = {arXiv},
  note = {\href{https://arxiv.org/abs/1406.4444}{arXiv~preprint~arXiv:1406.4444}},
}

@misc{prism_submodular,
  author = {Kothawade, Suraj and Kaushal, Vishal and Ramakrishnan, Ganesh and Bilmes, Jeff and Iyer, Rishabh},
  title = {{PRISM: A Rich Class of Parameterized Submodular Information Measures for Guided Subset Selection}},
  year = {2021},
  eprint = {2103.00128},
  archivePrefix = {arXiv},
  note = {\href{https://arxiv.org/abs/2103.00128}{arXiv~preprint~arXiv:2103.00128}},
}

@misc{prism_world,
  author = {Mirchev, Atanas and Kayalibay, Baris and Agha, Ahmed and van der Smagt, Patrick and Cremers, Daniel and Bayer, Justin},
  title = {{PRISM: Probabilistic Real-Time Inference in Spatial World Models}},
  year = {2022},
  eprint = {2212.02988},
  archivePrefix = {arXiv},
  note = {\href{https://arxiv.org/abs/2212.02988}{arXiv~preprint~arXiv:2212.02988}},
}

@misc{prism_pathology,
  author = {Shaikovski, George and Casson, Adam and Severson, Kristen and Zimmermann, Eric and Wang, Yi Kan and Kunz, Jeremy D. and Retamero, Juan A. and Oakley, Gerard and Klimstra, David and Kanan, Christopher and Hanna, Matthew and Zelechowski, Michal and Viret, Julian and Tenenholtz, Neil and Hall, James and Fusi, Nicolo and Yousfi, Razik and Hamilton, Peter and Moye, William A. and Vorontsov, Eugene and Liu, Siqi and Fuchs, Thomas J.},
  title = {{PRISM: A Multi-Modal Generative Foundation Model for Slide-Level Histopathology}},
  year = {2024},
  eprint = {2405.10254},
  archivePrefix = {arXiv},
  note = {\href{https://arxiv.org/abs/2405.10254}{arXiv~preprint~arXiv:2405.10254}},
}

@misc{prism_safety,
  author = {Neumann, Terrence and Jones, Bryan},
  title = {{PRISM: A Design Framework for Open-Source Foundation Model Safety}},
  year = {2024},
  eprint = {2406.10415},
  archivePrefix = {arXiv},
  note = {\href{https://arxiv.org/abs/2406.10415}{arXiv~preprint~arXiv:2406.10415}},
}

@misc{prism_imagination,
  author = {Wang, Yuhai and Xia, Jiawei and Sun, Bowen and Zhou, Rongxuan and Hu, Xiao and Shi, Yongliang and Du, Jing and Ye, Yang},
  title = {{PRISM: PRior-guided Imagination Sampling in world Models}},
  year = {2026},
  eprint = {2606.07974},
  archivePrefix = {arXiv},
  note = {\href{https://arxiv.org/abs/2606.07974}{arXiv~preprint~arXiv:2606.07974}},
}

@misc{prism_ehr,
  author = {Zhu, Yinghao and Wang, Zixiang and He, Long and Xie, Shiyun and Zheng, Xiaochen and Ma, Liantao and Pan, Chengwei},
  title = {{PRISM: Mitigating EHR Data Sparsity via Learning from Missing Feature Calibrated Prototype Patient Representations}},
  year = {2023},
  eprint = {2309.04160},
  archivePrefix = {arXiv},
  note = {\href{https://arxiv.org/abs/2309.04160}{arXiv~preprint~arXiv:2309.04160}},
}

@misc{prism_cell,
  author = {Moreira, Larissa Ferreira Rodrigues and Rodrigues, Leonardo Gabriel Ferreira and Moreira, Rodrigo and Backes, André Ricardo},
  title = {{PRISM: Perinuclear Ring-based Image Segmentation Method for Acute Lymphoblastic Leukemia Classification}},
  year = {2026},
  eprint = {2605.12851},
  archivePrefix = {arXiv},
  note = {\href{https://arxiv.org/abs/2605.12851}{arXiv~preprint~arXiv:2605.12851}},
}

@misc{trace_table,
  author = {Baek, Youngmin and Nam, Daehyun and Surh, Jaeheung and Shin, Seung and Kim, Seonghyeon},
  title = {{TRACE: Table Reconstruction Aligned to Corner and Edges}},
  year = {2023},
  eprint = {2305.00630},
  archivePrefix = {arXiv},
  note = {\href{https://arxiv.org/abs/2305.00630}{arXiv~preprint~arXiv:2305.00630}},
}

@misc{trace_attribution,
  author = {Wang, Cheng and Lu, Xinyang and Ng, See-Kiong and Low, Bryan Kian Hsiang},
  title = {{TRACE: TRansformer-based Attribution using Contrastive Embeddings in LLMs}},
  year = {2024},
  eprint = {2407.04981},
  archivePrefix = {arXiv},
  note = {\href{https://arxiv.org/abs/2407.04981}{arXiv~preprint~arXiv:2407.04981}},
}

@misc{trace_timeseries,
  author = {Li, Yuze and Zhu, Wei},
  title = {{TRACE: Time SeRies PArameter EffiCient FinE-tuning}},
  year = {2025},
  eprint = {2503.16991},
  archivePrefix = {arXiv},
  note = {\href{https://arxiv.org/abs/2503.16991}{arXiv~preprint~arXiv:2503.16991}},
}

@misc{trace_evaluation,
  author = {Chen, Yanyu and Jiang, Jiyue and Liu, Jiahong and Zhang, Yifei and Guo, Xiao and King, Irwin},
  title = {{TRACE: Trajectory-Aware Comprehensive Evaluation for Deep Research Agents}},
  year = {2026},
  eprint = {2602.21230},
  archivePrefix = {arXiv},
  note = {\href{https://arxiv.org/abs/2602.21230}{arXiv~preprint~arXiv:2602.21230}},
}

@misc{trace_monitor,
  author = {Mittapalli, Vijitha and Dani, Shreyaa Jayant and Pilli, Satya Srujana and Ansu, Snigdha and Teymoorianfard, Mohammadreza and Dernoncourt, Franck and Chen, Hongjie and Wang, Yu and Rossi, Ryan A. and Ahmed, Nesreen K.},
  title = {{TRACE: Trajectory Reasoning through Adaptive Cross-Step Evidence Aggregation for LLM Agents}},
  year = {2026},
  eprint = {2606.07054},
  archivePrefix = {arXiv},
  note = {\href{https://arxiv.org/abs/2606.07054}{arXiv~preprint~arXiv:2606.07054}},
}

@misc{sage_slam,
  author = {Liu, Xingtong and Li, Zhaoshuo and Ishii, Masaru and Hager, Gregory D. and Taylor, Russell H. and Unberath, Mathias},
  title = {{SAGE: SLAM with Appearance and Geometry Prior for Endoscopy}},
  year = {2022},
  eprint = {2202.09487},
  archivePrefix = {arXiv},
  note = {\href{https://arxiv.org/abs/2202.09487}{arXiv~preprint~arXiv:2202.09487}},
}

@misc{sage_attributes,
  author = {Nikolakopoulos, Athanasios N. and Kaul, Swati and Gade, Siva Karthik and Dubrov, Bella and Batur, Umit and Khan, Suleiman Ali},
  title = {{SAGE: Structured Attribute Value Generation for Billion-Scale Product Catalogs}},
  year = {2023},
  eprint = {2309.05920},
  archivePrefix = {arXiv},
  note = {\href{https://arxiv.org/abs/2309.05920}{arXiv~preprint~arXiv:2309.05920}},
}

@misc{sage_entropy,
  author = {Bonagiri, Vamshi Krishna and Vennam, Sreeram and Govil, Priyanshul and Kumaraguru, Ponnurangam and Gaur, Manas},
  title = {{SaGE: Evaluating Moral Consistency in Large Language Models}},
  year = {2024},
  eprint = {2402.13709},
  archivePrefix = {arXiv},
  note = {\href{https://arxiv.org/abs/2402.13709}{arXiv~preprint~arXiv:2402.13709}},
}

@misc{sage_search,
  author = {Badshah, Sher and Emami, Ali and Sajjad, Hassan},
  title = {{SAGE: A Search-AuGmented Evaluation of Large Language Models on Free-Form QA}},
  year = {2025},
  eprint = {2504.07385},
  archivePrefix = {arXiv},
  note = {\href{https://arxiv.org/abs/2504.07385}{arXiv~preprint~arXiv:2504.07385}},
}

@misc{sage_query,
  author = {Wang, Teng and Gong, Hailei and Zhang, Changwang and Wang, Jun},
  title = {{SAGE: Strategy-Adaptive Generation Engine for Query Rewriting}},
  year = {2025},
  eprint = {2506.19783},
  archivePrefix = {arXiv},
  note = {\href{https://arxiv.org/abs/2506.19783}{arXiv~preprint~arXiv:2506.19783}},
}

@misc{sage_audio,
  author = {Brigante, Francesco and Cerovaz, Luca and Marincione, Davide and Strano, Giorgio and Zhou, Luca and Rodolà, Emanuele and Mancusi, Michele},
  title = {{SAGE: Semantic Audio Generative Encoder}},
  year = {2026},
  eprint = {2609.32755},
  archivePrefix = {arXiv},
  note = {\href{https://arxiv.org/abs/2609.32755}{arXiv~preprint~arXiv:2609.32755}},
}

@misc{sageattention,
  author = {Zhang, Jintao and Wei, Jia and Huang, Haofeng and Zhang, Pengle and Zhu, Jun and Chen, Jianfei},
  title = {{SageAttention: Accurate 8-Bit Attention for Plug-and-play Inference Acceleration}},
  year = {2024},
  eprint = {2410.02367},
  archivePrefix = {arXiv},
  note = {\href{https://arxiv.org/abs/2410.02367}{arXiv~preprint~arXiv:2410.02367}},
}

@misc{sageattention2,
  author = {Zhang, Jintao and Huang, Haofeng and Zhang, Pengle and Wei, Jia and Zhu, Jun and Chen, Jianfei},
  title = {{SageAttention2: Efficient Attention with Thorough Outlier Smoothing and Per-thread INT4 Quantization}},
  year = {2024},
  eprint = {2411.10958},
  archivePrefix = {arXiv},
  note = {\href{https://arxiv.org/abs/2411.10958}{arXiv~preprint~arXiv:2411.10958}},
}

@misc{sageattention3,
  author = {Zhang, Jintao and Wei, Jia and Zhang, Pengle and Xu, Xiaoming and Huang, Haofeng and Wang, Haoxu and Jiang, Kai and Chen, Jianfei and Zhu, Jun},
  title = {{SageAttention3: Microscaling FP4 Attention for Inference and An Exploration of 8-Bit Training}},
  year = {2025},
  eprint = {2505.11594},
  archivePrefix = {arXiv},
  note = {\href{https://arxiv.org/abs/2505.11594}{arXiv~preprint~arXiv:2505.11594}},
}

@misc{census,
  author = {Parikh, Tapan},
  title = {{The One-Word Census: Answer-Choice Conformity Across 105 Language Models}},
  year = {2026},
  eprint = {2607.12796v4},
  archivePrefix = {arXiv},
  note = {\href{https://arxiv.org/abs/2607.12796v4}{arXiv~preprint~arXiv:2607.12796v4}},
}

@misc{ghost,
  author = {Brzozowski, Michał and Chung, Neo Christopher},
  title = {{The Ghost Couple: Correlated LLM Name Priors and Their Haunting of the Web and Academic Publishing}},
  year = {2026},
  eprint = {2606.02184v2},
  archivePrefix = {arXiv},
  note = {\href{https://arxiv.org/abs/2606.02184v2}{arXiv~preprint~arXiv:2606.02184v2}},
}

@misc{taste,
  author = {Parikh, Tapan},
  title = {{How Language Models Learn Taste: Brand Preference in LLMs}},
  year = {2026},
  eprint = {2610.28272v1},
  archivePrefix = {arXiv},
  note = {\href{https://arxiv.org/abs/2610.28272v1}{arXiv~preprint~arXiv:2610.28272v1}},
}

@misc{verbalized,
  author = {Zhang, Jiayi and Yu, Simon and Chong, Derek and Sicilia, Anthony and Tomz, Michael R. and Manning, Christopher D. and Shi, Weiyan},
  title = {{Verbalized Sampling: How to Mitigate Mode Collapse and Unlock LLM Diversity}},
  year = {2025},
  eprint = {2510.01171v4},
  archivePrefix = {arXiv},
  note = {\href{https://arxiv.org/abs/2510.01171v4}{arXiv~preprint~arXiv:2510.01171v4}},
}

@misc{hivemind,
  author = {Jiang, Liwei and Chai, Yuanjun and Li, Margaret and Liu, Mickel and Fok, Raymond and Dziri, Nouha and Tsvetkov, Yulia and Sap, Maarten and Albalak, Alon and Choi, Yejin},
  title = {{Artificial Hivemind: The Open-Ended Homogeneity of Language Models (and Beyond)}},
  year = {2025},
  eprint = {2510.22954v2},
  archivePrefix = {arXiv},
  note = {\href{https://arxiv.org/abs/2510.22954v2}{arXiv~preprint~arXiv:2510.22954v2}},
}

@misc{hivemind_review,
  author = {Schaeffer, Rylan and Miranda, Brando and Kazdan, Joshua and Chudnovsky, Jessica and Koyejo, Sanmi},
  title = {{How Strong Is the Evidence for the Artificial Hivemind? Reevaluating Evidence for the Open-Ended Homogeneity of Language Models}},
  year = {2026},
  eprint = {2609.33936v1},
  archivePrefix = {arXiv},
  note = {\href{https://arxiv.org/abs/2609.33936v1}{arXiv~preprint~arXiv:2609.33936v1}},
}

@misc{monoculture,
  author = {Thejaswi, Suhas and Kulshreshta, Juhi and Oettershagen, Lutz},
  title = {{Linguistic Monoculture in LLM-Assisted Language Use}},
  year = {2026},
  eprint = {2607.27134v1},
  archivePrefix = {arXiv},
  note = {\href{https://arxiv.org/abs/2607.27134v1}{arXiv~preprint~arXiv:2607.27134v1}},
}

@misc{acronym_tasks,
  author = {Veyseh, Amir Pouran Ben and Dernoncourt, Franck and Nguyen, Thien Huu and Chang, Walter and Celi, Leo Anthony},
  title = {{Acronym Identification and Disambiguation Shared Tasks for Scientific Document Understanding}},
  year = {2020},
  eprint = {2012.11760v4},
  archivePrefix = {arXiv},
  note = {\href{https://arxiv.org/abs/2012.11760v4}{arXiv~preprint~arXiv:2012.11760v4}},
}

@misc{llm_scientific_writing,
  author = {Liang, Weixin and Zhang, Yaohui and Wu, Zhengxuan and Lepp, Haley and Ji, Wenlong and Zhao, Xuandong and Cao, Hancheng and Liu, Sheng and He, Siyu and Huang, Zhi and Yang, Diyi and Potts, Christopher and Manning, Christopher D and Zou, James Y.},
  title = {{Mapping the Increasing Use of LLMs in Scientific Papers}},
  year = {2024},
  eprint = {2404.01268},
  archivePrefix = {arXiv},
  note = {\href{https://arxiv.org/abs/2404.01268}{arXiv~preprint~arXiv:2404.01268}},
}

@misc{sentence_bert,
  author = {Reimers, Nils and Gurevych, Iryna},
  title = {{Sentence-BERT: Sentence Embeddings using Siamese BERT-Networks}},
  year = {2019},
  eprint = {1908.10084},
  archivePrefix = {arXiv},
  note = {\href{https://arxiv.org/abs/1908.10084}{arXiv~preprint~arXiv:1908.10084}},
}

@misc{umap,
  author = {McInnes, Leland and Healy, John and Melville, James},
  title = {{UMAP: Uniform Manifold Approximation and Projection for Dimension Reduction}},
  year = {2018},
  eprint = {1802.03426},
  archivePrefix = {arXiv},
  note = {\href{https://arxiv.org/abs/1802.03426}{arXiv~preprint~arXiv:1802.03426}},
}

@misc{mpnet,
  author = {Song, Kaitao and Tan, Xu and Qin, Tao and Lu, Jianfeng and Liu, Tie-Yan},
  title = {{MPNet: Masked and Permuted Pre-training for Language Understanding}},
  year = {2020},
  eprint = {2004.09297},
  archivePrefix = {arXiv},
  note = {\href{https://arxiv.org/abs/2004.09297}{arXiv~preprint~arXiv:2004.09297}},
}
\endgroup
\clearpage
\onecolumn
\linespread{1}\selectfont
\small
\appendix
\section{Corpus definitions and sensitivity analyses}
\label{app:measurement}

\paragraph{Records and first versions.}
The primary population consists of \CorpusCount{} distinct arXiv identifiers with an eligible category tag and first submission in 2020--2026. The 2015--2019 extension contains \HistoricalCount{} records. Submission dates come from the first entry in the source version history. A single-version snapshot supplies original metadata directly; multi-version records receive explicit version-one metadata. All primary records have first-version titles and abstracts. Current titles and original titles are retained separately. The acquired artifact stores metadata, source links, and extraction evidence.

\begin{figure}[ht]
\centering\includegraphics{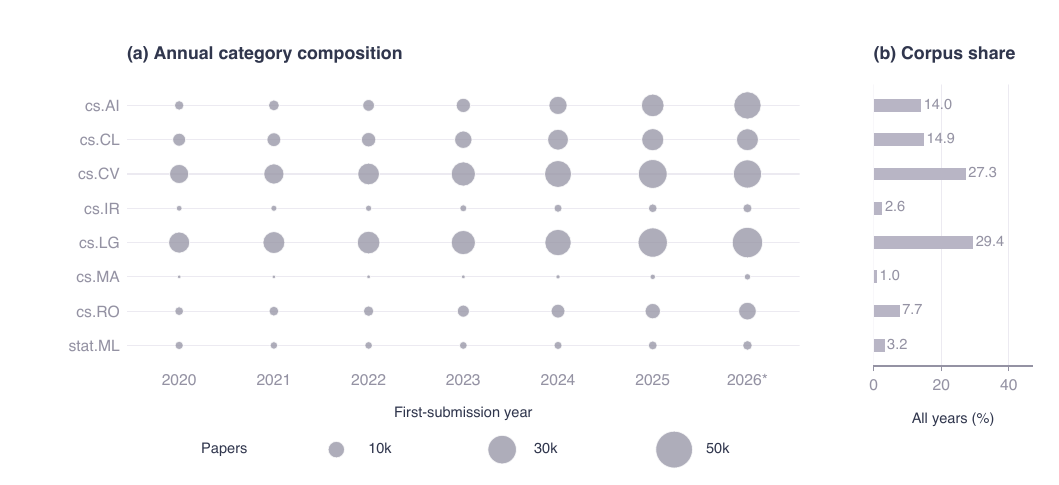}
\caption{\textbf{Corpus composition by category and year.} Each record is assigned to its first eligible category for this display. Category totals summarize the composition of the primary population. All primary records have first-version metadata.}
\label{fig:coverage}
\end{figure}

\paragraph{Title-label rules.}
The principal extractor matches a leading token with \verb|[A-Za-z][A-Za-z0-9+_.-]{1,39}|, followed by an ASCII or full-width colon and a nonempty title remainder. A frozen stop list removes generic leading words such as \emph{the}, \emph{towards}, \emph{learning}, \emph{survey}, and \emph{methods}. Unicode compatibility normalization and case folding define the comparison key. Punctuation and suffixes remain in the key. The format-filtered rule additionally removes \emph{position}, \emph{sok}, \emph{demo}, \emph{review}, \emph{tutorial}, and \emph{perspective}. The capitalization filter retains 3--20-character labels with an uppercase letter after the first. Case-sensitive matching retains the \mbox{original label spelling}.

\paragraph{Reference periods and head removal.}
The pooled baseline combines all labelled records from 2020, 2021, and 2022 before applying Equation~\ref{eq:corpus_collision}. A second reference divides the sum of matching pairs within those years by the sum of all eligible pairs within the same years. Their all-label 2026 ratios are 4.5199 and 4.5157, respectively. Against 2020 alone the ratio is 3.74; against 2022 it is 4.81. Truncating every cohort to January 1 through October 27 gives a 4.83 ratio for the 2022 comparison. The exported sensitivities retain all of these reference definitions.

The fixed head list ranks non-format labels by total frequency across 2020--2026. Removing the first $k$ names applies the identical list to the reference and 2026 populations, and recomputes both matching and eligible pair counts. Figure~\ref{fig:namespace} reports $k=0,1,5,10,20$. The fixed-category estimate uses each paper's first eligible category and weights the categories by their labelled 2020--2022 population. The annual data retain every extraction rule for initial and latest titles.

\paragraph{Author-disjoint pairs.}
Author names are normalized with Unicode compatibility normalization, case folding, and whitespace normalization. Cornell's parsed surname, given-name, and suffix fields are reassembled into display order before comparison with Atom names. Each within-cohort paper pair sharing at least one normalized full-name string is removed from both the matching-pair numerator and eligible-pair denominator. Each pair is removed once even if it shares several names. This operation measures author-string overlap. The original author fields remain available for identity-aware analysis. Source-declared method lineages and detailed case records provide separate evidence about reuse of a named method.

\paragraph{Semantic topic partitions.}
Topic analyses use the 144,523 non-format labelled records in the title-retrieval collection. Each abstract is masked at occurrences of its own title label before encoding. Normalized 384-dimensional all-MiniLM-L6-v2 embeddings provide the representation \citep{sentence_bert,minilm_model}. The model encodes up to 256 wordpieces. MiniBatchKMeans uses 100, 200, or 400 clusters with seed 20261029, batch size 2,048, three initializations, initialization sample size 12,000, maximum 100 iterations, and a 15-step no-improvement criterion. Each resolution is fitted once across the common collection and reused for every year. Within-topic collision counts matching labels among same-cluster pairs; cross-topic collision counts matching labels among different-cluster pairs. Each numerator is divided by the corresponding eligible-pair count. Cluster assignments, centroids, and aggregate counts are retained. These groups operationalize semantic similarity of abstracts; method lineages are identified from source evidence.

The joint sensitivity uses the 200-cluster partition and removes cross-topic pairs sharing any normalized author name. The pooled baseline contains 2,085 matches among 313,225,897 eligible pairs; 2026 contains 27,080 among 925,495,735. Dividing the two conditional collision rates gives 4.3957. Pooling only same-year pairs from the three reference years gives a 4.5190 ratio. This analysis uses the same author normalization and pair-deduplication rule as the author-only sensitivity.

For the 200-cluster partition, the within-year reference gives cross-topic growth of 4.43 and within-topic growth of 0.88. These pair-restricted references accompany the pooled-record estimates in the exported analysis.

\paragraph{Name-group uncertainty.}
For each comparison, labels are drawn with replacement from the union of the two periods. A draw copies the complete name group, retaining its record count and topic counts. Duplicate draws are treated as separate resampled groups. Matching-pair counts therefore receive linear group multiplicities, and eligible pairs are recalculated from the resampled record and topic totals. The same weights apply to both periods. Percentile intervals use 2,000 draws; all reported comparisons have valid denominators in every draw. This procedure quantifies sensitivity to the observed collection of name groups.

\paragraph{A second sentence encoder.}
The independent representation check uses all-mpnet-base-v2 \citep{mpnet,mpnet_model}, a different encoder architecture with 768-dimensional sentence vectors. It uses its documented 384-wordpiece limit, compared with 256 for the MiniLM configuration. Both encode abstracts after masking the record's own title label. The model revision is \texttt{e8c3b32edf5434bc2275fc9bab85f82640a19130}. MiniBatchKMeans uses 200 clusters, seed 20261030, batch size 2,048, three initializations, and at most 100 iterations. Cached half-precision vectors are renormalized in full precision for clustering. The exported configuration records the complete encoding and partition settings. The check changes the encoder representation while retaining the paper population and outcome definitions.

The two partitions agree on the cross-topic status of 114,028 same-name pairs and the same-topic status of 963 pairs. MiniLM alone classifies 1,099 pairs as same-topic; MPNet alone classifies 1,139. Same-topic pair Jaccard similarity, the intersection divided by the union, is 0.301. Across all records, adjusted Rand agreement is 0.318 and normalized mutual information is 0.642. These partition measures describe shared group membership after accounting for the arbitrary cluster labels. The historical ratios and fixed-size retrieval comparisons reproduce the central pattern across these different boundaries.

\paragraph{Projection for illustration.}
The shared UMAP display \citep{umap} includes 20,000 sampled records outside the eight most frequent name groups and all 659 records belonging to those groups. Every name panel uses this common fit and coordinate system. The input is the own-label-masked MiniLM representation. UMAP uses cosine distance, 30 neighbors, minimum distance 0.18, 300 epochs, and seed 20261030. The gray background is a vector density aggregation. The projection illustrates distributions; topic counts and statistical comparisons use the original representations and frozen cluster assignments.

\paragraph{Name-level topic breadth.}
The topic degree of a name is the number of semantic clusters containing its records. Raw degree grows with the number of observed uses, so the artifact also reports expected topic coverage in a sample of five records drawn without replacement. For a name with $n$ records and cluster counts $n_k$, this quantity is $\sum_k[1-\binom{n-n_k}{5}/\binom{n}{5}]$. The 18 labels eligible in both periods give the fixed-volume comparison reported in the main text. The exported measurements retain raw topic degree and this normalized coverage for every eligible name.

\subsection{Source-linked name cases}
\label{app:cases}

\paragraph{Additional PRISM sources.}
Pathology encoding and world-model planning provide further research objects \citep{prism_pathology,prism_imagination}. A patient-representation paper illustrates a label added after the original submission \citep{prism_ehr}. Earlier person re-identification work supplies a 2014 historical example \citep{prism_2014}. The exported records retain all three observed exact-label spellings and the separately counted suffix variants.

\paragraph{Additional TRACE sources.}
Agent monitoring supplies a further process-oriented use \citep{trace_monitor}. The source-linked group retains original titles, current titles, and explicit capitalization. Separator suffixes and concatenated extensions remain separate from its 120-record exact-label group.

\paragraph{Additional SAGE sources.}
Product-attribute generation \citep{sage_attributes}, search evaluation \citep{sage_search}, and query rewriting \citep{sage_query} extend the source-linked task coverage. Separately labelled SageAttention successors retain their source-declared lineage links. These records support inspecting query ambiguity \mbox{and method relationships.}

\section{Supporting model-naming case study}
\label{app:protocols}

\paragraph{Controlled inputs.}
The frozen task file contains 360 distinct four-component specifications drawn from 16 component types, with component identities, licensed phrase realizations, and six research contexts. The original rephrasing changes phrase realization and reverses list order. The crossed experiment below separates those factors while retaining the confirmation families.

\subsection{Crossed description experiment}
\label{app:factorial}

The design uses the same 270 controlled confirmation specifications. Each specification crosses phrase realization 0 versus 1, original versus reversed list order, and the original base versus rephrased description frame. The first frame begins with a proposed method that combines the component list. The second introduces a proposed system and lists its required components. Each frame retains its complete instruction wording, including the statement that order is immaterial. Phrase and order vary independently within each frame. The frozen protocol contains every rendered description and all factor labels before requests are enqueued.

Each of the eight cells contains four fresh requests per family, giving 1,080 requests per cell and 8,640 overall. A fixed seed shuffles the complete assignment list before dispatch. DeepSeek Flash uses temperature 1.0, thinking disabled, a 512-token cap, and JSON responses through the existing execution and accounting pipeline. All 8,640 assignments finish; 8,590 yield syntactically valid names. The 50 invalid outputs retain their assigned-request denominators.

The two outcomes are the fraction of assignments selecting the exact normalized name TRACE and cross-family name collision. Each factor effect averages the four matched contrasts defined by the other factors. Intervals use 1,000 common family-bootstrap draws. The 100,000 permutation draws swap the complete two-level factor profiles within each family, preserving the other factor settings. Plus-one permutation values receive Holm adjustment across the six factor--outcome comparisons. For TRACE, adjusted values are 0.0111 for frame, 1.0 for phrase, and 0.000060 for order. For collision, frame gives 121.68 [57.82, 205.46] additional matches per 100,000 pairs with adjusted $p=0.00012$; phrase gives $-18.24$ [$-86.24$, 60.26] with $p=1$; order gives 182.78 [73.57, 315.51] with $p=0.000060$.

The trajectory analysis is exploratory. Sixty-two families contain the trajectory-memory component. Reversal places it first in 54 of those families; score aggregation occupies the first position in the remaining eight. The paired first-position comparison retains the same 54 families under both orders and averages across phrase and frame settings. It contains 21 TRACE outputs among 864 forward assignments and 185 among 864 reversed assignments. Source expansions and the 20-family surface-word audit are retained separately from this controlled order contrast.

\begin{figure}[ht]
\centering\includegraphics{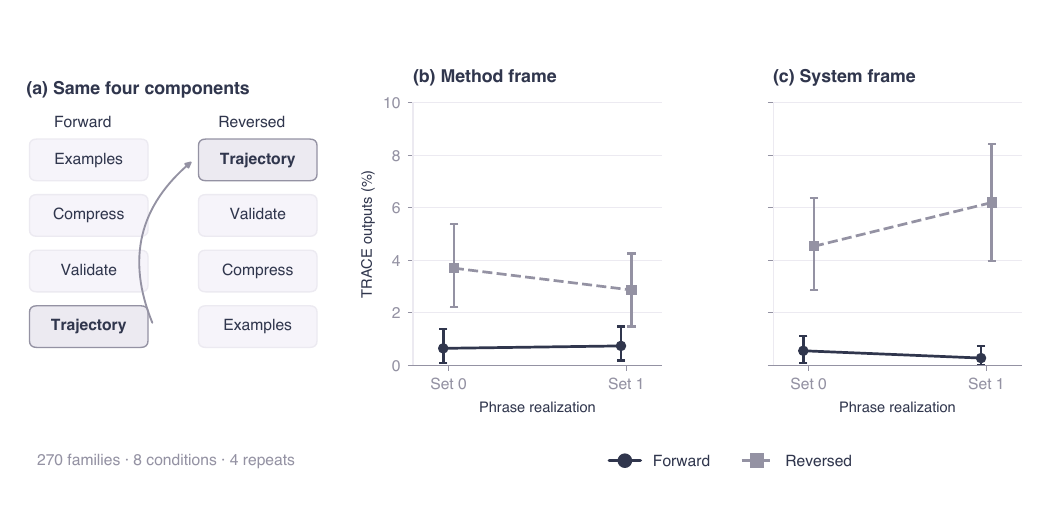}
\caption{\textbf{List order selects a particular name attractor.} The same components are presented in forward and reversed order under two phrase sets and two frames. TRACE frequency uses all assigned requests; intervals resample families. The component example identifies the trajectory-memory position change.}
\end{figure}

\subsection{Original comparison materials}

\paragraph{Natural inputs and matched author names.}
The 360 arXiv abstracts were frozen from available metadata at task construction. Case-insensitive word-boundary matching replaces detected names with a generic method reference. Associated expansions and technical terms remain in the source text. Every task retains its paper identifier and current verified submission date. Of the 270 confirmation abstracts, 220 source papers have an extractable current title label. The direct author--model comparison retains these same 220 families in both distributions. Author naming contributes one observed label per family; model naming contributes the 12 assigned outputs for that family. The 50 other abstracts continue to enter the all-abstract model estimate. Across all 270 confirmation abstracts, DeepSeek collision is 0.306 per 100,000 pairs. On the 220 matched families, the observed author labels have zero matching pairs among 24,090 possible pairs, and model collision is 0.115 per 100,000; the paired interval is [0, 0.521]. These counts describe the frozen matched sample.

\paragraph{Requests and validity.}
Controlled naming has three conditions, and abstract naming has one, each with 12 requests per family. The DeepSeek free-naming allocation contains 17,280 requests, including 31 terminal failures. Syntactic name validity requires a nonempty single-line name of at most 80 characters in the expected JSON object. DeepSeek Flash uses temperature 1.0, thinking disabled, a 512-token cap, and a JSON-object response. Retries are recorded as attempts of the original assignment. GPT-6-Astra uses isolated ephemeral sessions with a tool-free instruction, event-level tool-use checks, and low reasoning effort. That interface manages temperature and output limits. Its 360 assignments cover matched controlled and abstract inputs plus controlled rephrasing. The comparison identifies each model together with its actual access configuration. On the matched 60 controlled families, the original GPT estimates are 84.75 and 14.12 collisions per 100,000 assigned pairs under base and rephrased descriptions; the paired interval spans zero. The corresponding DeepSeek values are 169.49 and 353.11. Each observation retains its recorded access configuration.

\paragraph{Attractor decomposition.}
A name's collision contribution is the mean of $\widehat p_i(a)\widehat p_j(a)$ over distinct families, where $\widehat p_i(a)$ is its assigned-request frequency in family $i$. Summing names recovers total collision. Removing TRACE sets its contribution to zero in both conditions. Common family-bootstrap weights retain the paired comparison and assigned-request denominators.

\paragraph{Family uncertainty.}
Let $w_i$ be the number of copies of family $i$ in a bootstrap sample. The weighted estimator is
\begin{equation}
\widehat\C(w)=\frac{\sum_a(\sum_i w_i\widehat p_i(a))^2-\sum_i w_i^2\sum_a\widehat p_i(a)^2}{(\sum_i w_i)^2-\sum_i w_i^2}.
\end{equation}
The subtraction excludes copies of the same original family. Paired contrasts share 1,000 bootstrap draws. Two-sided permutation values use $(b+1)/(B+1)$ with $B=100{,}000$ complete within-family swaps. Holm adjustment \citep{holm} applies to the 16 original contrasts and, separately, to the six factorial contrasts defined above. The exported results retain each comparison family and unadjusted value.

\paragraph{Expansion-word audit.}
A fixed seed samples 20 of the 37 original confirmation families producing TRACE. The first stored TRACE response per sampled family is matched against a frozen dictionary of component words. Six expansions contain cues for all four component types and 14 have at least one unmatched type. This is a surface-mention count. The dictionary, inputs, expansions, and matched cues are exported together, so paraphrases outside the dictionary remain inspectable.

\section{Grammar and lexical references}
\label{app:selection}

\paragraph{Equal-family comparison.}
On the flexible supports, the model's compliant-name collision is 125.16 per 100,000 equal-weight family pairs, with interval [62.21, 239.73]. The attested-token and frequency-weighted references give 521.33 and 1,631.69. Model/reference ratios are 0.240 [0.118, 0.439] and 0.0767 [0.0400, 0.1350]. The uniform legal reference gives 7.93. These comparisons locate the model between the structural and lexical sampling references.

\begin{figure}[ht]
\centering\includegraphics{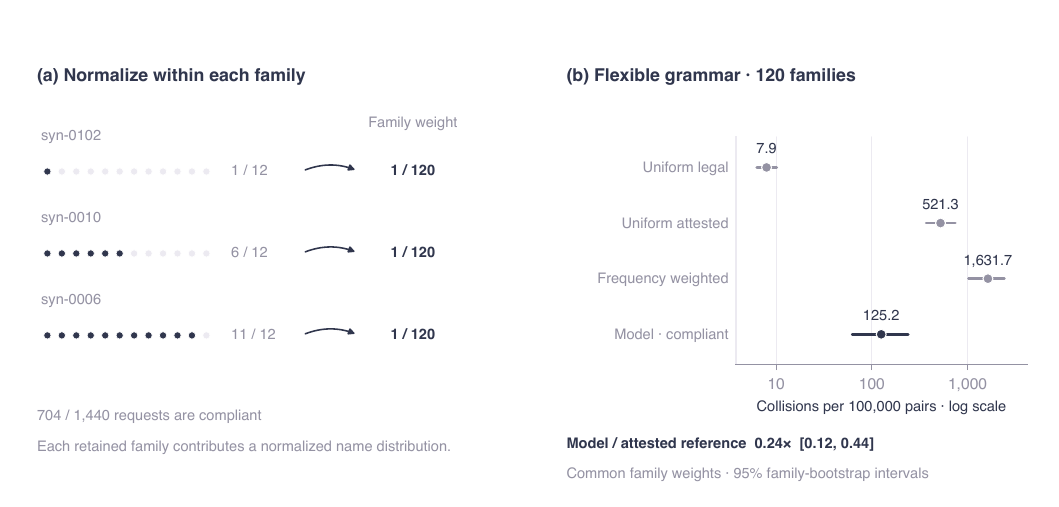}
\caption{\textbf{Reference distributions determine the comparison scale.} Each model family is normalized over its compliant outputs, and every family has the same weight as in each reference. The support and family sets are shared.}
\end{figure}

\paragraph{Feasibility and execution.}
An expansion is compliant when it uses exactly one licensed phrase for each required component, obeys the condition's order rule, and forms the requested name from initials. Restricted supports contain 8--16 distinct names; flexible supports contain 76--816. The 479 unique family--target pairs have four repeats under both rules, giving 3,832 requests. The same targets appear in both conditions. There are 532 feasible restricted requests and 1,916 feasible flexible requests. Their successful counts are 512 and 1,397. On the common set of 532 requests, the successful counts are 512 and 444. Success on feasible restricted targets is 96.2\%, compared with 72.9\% on feasible flexible targets. Within the shared feasible subset, flexible execution succeeds on 83.5\% of requests. These comparisons separate target coverage from execution on available targets.

\paragraph{Legal and lexical choice baselines.}
The complete-support experiment lists every distinct legal name once in randomized order and asks for a memorable name with a valid expansion. It allocates 12 requests per family and rule. Uniform legal choice places equal mass on every listed name. A second structural baseline samples licensed phrase-and-order paths, preserving expansion multiplicities. Lexical baselines use the English large wordlist in wordfreq 3.1.1 \citep{wordfreq}. Frequency-weighted choice assigns weight $\max(f(a),10^{-8})$ to each legal name. Uniform attested-token choice assigns equal mass to legal names with $f(a)\ge10^{-6}$; a family with an empty qualifying set uses its full legal support. These token frequencies include attested abbreviations and proper names. The artifact records support and fallback counts for every family.

Collision conditional on both outputs being compliant divides joint compliant-name collision by the probability that both members of a distinct-family request pair comply. This conditions the original request-pair distribution and reweights family pairs accordingly. The primary model--reference comparison normalizes compliant outputs within each family and then weights all distinct-family pairs equally. Both rules have at least one compliant output in every family. Under flexibility, joint collision is 40.56 per 100,000 assigned pairs, request-pair conditional collision is 169.89, and equal-family conditional collision is 125.16.

\paragraph{Fixed menus and selection.}
The six-name control chooses a fixed subset of each restricted support and supplies valid example expansions under both rules. All six names are feasible in both conditions. Joint collision rates are 35.01 and 25.39 per 100,000 assigned pairs under restriction and flexibility. The paired difference interval spans zero.

The complete original comparison set, candidate banks, and selection outcomes are retained in the data release. The manuscript reports the controls used to interpret the current naming case study.

\FloatBarrier
\section{Retrieval protocol and reproducible artifacts}
\label{app:retrieval}
\paragraph{E5 configurations.}
The dense extension uses \texttt{intfloat/e5-base-v2} at revision \texttt{f52bf8ec}, with the complete revision identifier in the artifact. Documents receive the \texttt{passage:} prefix and queries receive \texttt{query:}, following the model instructions. The model uses 768-dimensional attention-masked mean pooling and a 512-wordpiece limit. Encoding uses half precision; pooling and normalization use full precision. Cached half-precision vectors are renormalized and scored in full precision with tie tolerance $10^{-6}$. The title configuration runs all 8,400 frozen queries in both candidate scopes. The document-field extension uses title plus abstract for the 2,800 contextual queries, retaining the same source terms and targets. Its results quantify retrieval when the indexed document includes the source of the query clues.

\paragraph{Listwise language-model ranking.}
DeepSeek-Flash receives the query and all titles in the exact normalized-name group. Candidate order uses a seeded permutation independent of the target's relevance. Each candidate has an opaque identifier, and the requested JSON ordering must contain every identifier exactly once. The temperature is zero, thinking is disabled, and the output cap is 2,048 tokens. The fixed protocol assigns 1,400 shared-name contextual queries. A 20-query execution check precedes the remaining requests; all 20 remain in the final denominator. Network retries follow the existing transient-error policy, with at most three attempts. Invalid rankings receive zero credit and trigger no quality retry. The final records contain 1,398 valid rankings and two invalid outputs.

The API returns the alias \texttt{deepseek-flash}; the recorded system fingerprint is retained with the responses. The official service documentation described that alias as DeepSeek-V4.1-Flash-1017 at execution. The 1,400 requests add USD~0.193195 under the project's conservative token accounting. The artifact exports the complete candidate inputs, response orderings, usage, finish reasons, and validation states. The standalone evaluator reproduces all reported aggregate metrics from the submitted rankings.

\paragraph{Targets and queries.}
The 144,523-title index contains non-format title labels. The shared-name sample selects 1,400 of 28,152 eligible records with an abstract of at least 100 characters. Unique controls match without replacement on year, first eligible category, and title-token-length interval, using fewer than 10, 10--14, 15--19, and at least 20 tokens. All matches are exact on these criteria and frozen before scoring. Abstract terms are lowercase alphabetic tokens of at least four characters, ranked by $(1+\log\mathrm{tf})\log((N+1)/(\mathrm{df}+1))$ using term-frequency and inverse-document-frequency weighting \citep{tfidf}. The target label and a fixed stop list are removed. Document frequencies come from the full abstract collection. The same term-selection rule applies to \mbox{every name group.}

\paragraph{Scores and inference.}
BM25 uses $k_1=1.2$, $b=0.75$, and $\log(1+(N-\mathrm{df}+0.5)/(\mathrm{df}+0.5))$ inverse document frequency. Candidate filtering retains full-index statistics. MiniLM uses normalized 384-dimensional vectors and its 256-wordpiece limit. With $h$ higher scores and $t$ ties, expected reciprocal rank is $t^{-1}\sum_{j=1}^{t}(h+j)^{-1}$; Top-1 and Top-10 use analogous tie-averaged inclusion probabilities. Tie tolerances are $10^{-10}$ for BM25 and $10^{-6}$ for MiniLM.

Two thousand bootstrap draws resample complete name groups, retaining their targets and all query conditions. Paired differences use common weights. Group sizes are 1, 2, 3--4, 5--9, 10--19, and at least 20. Unique-name comparisons are metadata-matched observations; contextual reranking fixes each target's candidates. The release retains all 36,400 original-baseline outcomes, matched pairs, queries, scores, rank bounds, leading identifiers, and the model revision. The extensions add 23,800 outcomes, giving 60,200 in total.

\paragraph{Topic-localized retrieval outcomes.}
The added analysis reuses every shared-name target, its exact-name candidates, the full-index BM25 statistics, and the cached dense vectors. At each of 100, 200, and 400 clusters, the first-rank tie set is split into the target, other records in its topic, and records in other topics. The three tie-averaged probabilities sum to one for every query. The gain from adding content terms equals the reduction in same-topic error plus the reduction in cross-topic error. The cross-topic share divides the second reduction by the total gain. This ratio is a signed contribution to a net change and can exceed one when same-topic error increases. The 2,000 bootstrap draws resample entire target-name groups and preserve both query conditions. The two-record comparison fixes candidate count and distinguishes the topic of the single competitor. Its groups contain 481 cross-topic and 59 same-topic competitors and support a descriptive comparison.

\section{Lexical classes and writing-style association}
\label{app:style_proxy}

\paragraph{Fixed lexical-frequency bands.}
The normalized alphabetic title labels receive their English large-wordlist frequency from wordfreq 3.1.1 \citep{wordfreq}. Positive frequencies $f$ map to Zipf values $z=\log_{10}f+9$. The six fixed bands are unavailable or nonalphabetic, positive-frequency $z<2$, $[2,3)$, $[3,4)$, $[4,5)$, and $[5,\infty)$. The bands were specified before calculating the profile and all six are retained. The lexical resource contains words, abbreviations, and proper names. Unavailable frequencies occupy a separate plotted position; their estimates have no connecting segment to the ordered positive-frequency bands.

For each period and band, record share divides its record count by the population count; matching-pair share divides its cross-topic same-name pair count by all cross-topic same-name pairs. Conditional collision divides the band's matching pairs by all different-topic record pairs inside that band. Its growth ratio compares 2026 with pooled 2020--2022. A band's contribution to the population rate instead uses the full population's cross-topic pair denominator. Changes in these additive contributions define its share of net growth. Summing Zipf $[3,5)$ gives 12.4\% of 2026 records and 89.4\% of cross-topic matching pairs, with name-group intervals [11.5, 13.4] and [81.4, 93.6]. Its net-growth contribution is 98.0\% [92.7, 101.2]; the signed contribution may exceed 100\% when other bands decline.

Two thousand paired draws resample complete normalized name groups from the union of the periods. The same multiplicities apply across periods and frequency bands. Duplicate sampled groups receive distinct resampled keys, and denominators are recomputed from copied record and topic counts. Every band has 2,000 finite growth estimates. An independent explicit-copy calculation verifies the point counts and 100 small resampled cases. Combining the bands at Zipf three reconstructs the earlier $10^{-6}$ class shares within $10^{-12}$.

The name-level scatter displays all 8,190 labels used by at least two records. Point area is proportional to record count, and vertical position counts occupied topics in the fixed 200-cluster partition. The twenty labelled names are selected by full-corpus frequency, with alphabetical tie breaking. Singleton labels remain in all frequency-band calculations. Overlapping growth intervals retain uncertainty in comparisons between the bands.

\paragraph{Binary attestation sensitivity.}
The primary class requires a normalized alphabetic label with wordfreq English frequency at least $10^{-6}$. The supplementary thresholds are $10^{-7}$ and $10^{-5}$. Each label has one fixed class across all years. Cross-topic collision within a class uses that class's eligible pairs. Its contribution to the population rate divides its matching pairs by the population's cross-topic denominator. Contributions therefore sum to the population rate. At $10^{-7}$, the attested class contains 17.9\% of 2026 records and 97.2\% of cross-topic matching pairs. At $10^{-5}$, it contains 5.6\% of records and 40.5\% of matching pairs, with conditional growth of 4.03 for that class and 4.10 for the remaining labels. The stricter boundary partitions the growing vocabulary across both classes. The fixed primary threshold matches the previously specified lexical reference; all thresholds and yearly counts are exported.

\paragraph{Direction of the writing-style association.}
Higher affinity to the published LLM-associated vocabulary is associated with fewer cross-topic namesakes under year, topic, and abstract-length controls. The 2025--2026 odds ratio is 0.87 [0.84, 0.91]. Adding label-attestation class gives 0.96 [0.90, 1.03], with an interval spanning one. These estimates report a negative association under the first specification and a near-null association under the fuller specification. The score measures reference-word affinity in original abstracts. Actual author use of LLMs is unobserved, so the analysis identifies a writing-style association.

\paragraph{Published-reference style score.}
The reference vocabulary is the 8,191-word Computer Science distribution released with \citet{llm_scientific_writing}. Each word receives its published AI-versus-human log-odds difference. For each original abstract, we mask its own title label, extract unique reference-vocabulary tokens, and average their weights. The analysis retains 138,132 records whose original title has the same normalized label and whose abstract contains at least 20 reference tokens. This is a document-level lexical-affinity adaptation. The published estimator fits a population mixture to sentence-level likelihoods.

The outcome is whether a record has a same-year namesake in another topic among the retained records. High and low scores are split at the median within each matching stratum. The analyses successively match year and topic, add abstract length, and add label-attestation class. Length intervals are below 120, 120--199, and at least 200 tokens. Strata require at least 20 records and at least five in each score group. Mantel--Haenszel odds ratios and standardized risk differences are accompanied by 1,000 paired name-group bootstrap draws. The fully stratified analysis retains 111,047 records overall and 68,158 in 2025--2026. The full-period odds ratio is 0.95 [0.89, 1.01]; the 2020--2022 ratio is 1.00 [0.78, 1.33]. The artifact retains all period and specification results, source weights, and provenance.

\end{document}